\documentstyle[11pt]{article}
\title{ Temporal Modulation of Traveling Waves in the Flow Between
Rotating Cylinders With Broken Azimuthal Symmetry}
\author{ Sarath G. K. Tennakoon and C. David  Andereck\\
{\em Department of Physics, The Ohio State University}\\
{\em 174, W18th Avenue,Columbus, Ohio  43210}\\
John. J. Hegseth\\
{\em Department of Physics, University of New Orleans}\\
{\em New Orleans, Louisiana 70148}\\
Hermann Riecke\\
{\em Department of Engineering Sciences and Applied Mathematics}\\
{\em Northwestern University}\\
{\em Evanston, Illinois 60208}}
\newcommand{\be}{\begin{equation}}
\newcommand{\ee}{\end{equation}}
\newcommand{\bea}{\begin{eqnarray}}
\newcommand{\eea}{\end{eqnarray}}

\date{\today}
\begin{document}
\pagebreak
\maketitle

\begin{abstract}

The effect of temporal modulation on traveling waves in the flows in two
distinct systems of rotating cylinders, both with broken azimuthal symmetry, has
been investigated.  It is shown that by modulating the control parameter at 
twice the critical frequency one can excite phase-locked standing waves and
standing-wave-like states which are not 
allowed when the system is rotationally symmetric.  We also show how previous 
theoretical results can be extended to handle patterns such as these, that 
are periodic in two spatial direction. 
\end{abstract}


\pagebreak
 
\section{ Introduction }
The behavior of pattern forming systems under external time-periodic forcing 
has been the subject of intense investigation over the past several 
years\cite{Davis,Niemela}.  Since such periodically
forced flows are common in nature and technological
applications\cite{SemiHa76}, a knowledge
of their stability properties may have important practical applications.  

There have been several experimental and theoretical studies investigating the
effect of time-periodic forcing on the 
stability of the axisymmetric {\em steady}
spatial patterns produced in the flow between two rotating
cylinders (Taylor-Couette flow) 
\cite{Don64,Ahlers85,WalDon88,Hall75,Riely76,Baren89,Car81}. 
These studies found that temporal modulation of control parameters can
stabilize or destabilize the primary flows; the spatial properties of
the base flow and the axisymmetric Taylor vortex flow remain, however,
 largely unaffected by the modulation.  Only a very small shift of
the instabilities in the parameter space has been observed in these systems. 

For wave structures, however, recent studies by Riecke {\em et al.}
\cite{RiCr88} and
Walgraef\cite{Wa88} have shown that for the appropriate frequency of an
external forcing a strong
resonance may occur between the forcing and the spatio-temporal
patterns. This can transform initially stable traveling 
waves into standing waves or
quasi-periodic structures.  This kind of behaviour has been experimentally
observed in electroconvection of nematic liquid crystals and in binary fluid
convection by 
Rehberg {\em et al.} \cite{rehberg}. It is also related to the parametric 
excitation 
of surface waves (Faraday waves) \cite{cross}.

In this paper we investigate the effect of temporal modulation on the traveling
wave patterns that appear near the primary bifurcation boundary of two rotating
cylinder systems with broken rotational symmetry.  First, we will discuss the
theoretical background for this study. 
Next, we will describe our experimental studies of the modulated 
{\em Taylor-Dean} and {\em Taylor-Couette} systems. Finally, we will 
compare the experimental observations with the theoretical predictions. 

\section{Theory}
 
Using symmetry arguments and suitable amplitude equations it has been shown
previously \cite{RiCr88,Wa88,Ri90a} that a resonant temporal forcing can 
excite standing waves in an effectively one-dimensional extended system 
undergoing a Hopf bifurcation to 
traveling waves. Here we will extend those arguments to systems which 
are periodic
in two directions. This is necessary for spiral vortex flow
since the azimuthal symmetry of the 
concentric Taylor system imposes additional constraints on the amplitude 
equations.

The excitation of standing waves by resonant forcing can be understood 
intuitively
by an extension of the standard example of a parametrically forced pendulum. 
In that case, a periodic parametric forcing such as, e.g., 
the periodic variation of the length of the
pendulum, leads to a periodic pumping of energy in and out of the pendulum. 
In order to effect a net increase in the pendulum's energy the 
pumping and the pendulum have to have a certain phase relationship; 
in particular, the frequencies have to be resonant. 
Maximal efficiency is achieved if the forcing frequency is twice the natural 
frequency of the pendulum (e.g., decreasing the length of the pendulum 
whenever it travels downward).

Considering extended systems as an array of coupled pendula makes it clear
that only standing waves but not traveling waves can be excited parametrically: 
in a standing wave all oscillators have the same phase (albeit different 
amplitudes), and
therefore all of them can satisfy the phase relationship necessary for 
excitation
simultaneously; 
in a traveling wave the phases are distributed evenly over the circle and the 
out-of-phase oscillators will lose the energy gained by the 
in-phase-oscillators. 

To consider spiral vortex flow, which arises in a fluid between
counter-rotating concentric cylinders\cite{andereck}, one has 
to extend this consideration to two dimensions, since 
the spirals are periodic along the axis {\it and} along the azimuthal direction.
Therefore, the superposition of oppositely traveling spirals (`ribbons') 
leads to waves which
are standing waves along the axis but traveling waves in the azimuthal 
direction.
Thus, the periodic forcing may pump energy into the ribbon at some azimuthal 
angle but will extract energy at the same time at a different angle.  For a 
net increase
these energies have to be different, which is achieved by an azimuthally 
dependent perturbation of the forcing. Alternatively, the azimuthal symmetry
of the system itself can be broken.
Maximal efficiency is achieved, in analogy to the temporal forcing, if the 
azimuthal wave number of the perturbation is twice the azimuthal wave number 
of the ribbon. 

 The above arguments can be made precise by considering spiral vortex flow 
near its onset where its amplitude is small. Then the periodic forcing as 
well as the azimuthal
perturbation can also be taken small and the system can be described by
coupled amplitude equations. The crucial ingredients are the critical 
eigenvectors of the system, i.e. those modes which have small (or zero) 
growth rate, since
they define the center manifold of the system in this parameter regime. 
In the present
case the critical modes are the two spiral wave modes of the flow field 
${\bf v}$ 
as well as the periodic forcing $F$ and the azimuthal perturbation $P$,
\bea
{\bf v} &=&A(T) \,e^{iqz+i\omega_h t+im\phi} \,{\bf f}_a(r)+B(T) 
\,e^{iqz-i\omega_h 
t-im\phi} \,{\bf f}_b(r)+c.c.+h.o.t., \label{e:v}\\
F&=& b e^{i\omega_e t} + c.c., \label{e:F}\\
P &=& w e^{i n \phi} + c.c. \label{e:P}
\eea
Here $T$ is a slow time and ${\bf f}_{a,b}(r)$ are the radial eigenvectors of 
the spiral vortex flow.
The axis of the cylinders is taken along the $z$-direction and $\phi$ is the 
azimuthal
angle.
The frequencies $\omega_h$ and $\omega_e$ are the Hopf frequency of the spirals
and the frequency of the forcing, respectively. Below we will choose 
$\omega_e$ close
to $2 \omega_h$. The difference will determine the detuning.
Although the forcing and the
azimuthal perturbation are imposed externally we consider them as 
dynamical variables and include them in the center manifold. 
This simplifies the derivation
of the relevant amplitude equations for $A(T)$ and $B(T)$. Their form is 
obtained by
considering the most general polynomial which is consistent with the symmetries
of the system (in the absence of forcing, $F=0$, and the azimuthal 
perturbation $P=0$): translation symmetry in
space along $z$, rotational symmetry in $\phi$, translational symmetry
in time $t$, and reflection symmetry in $z$.
These symmetries allow a linear coupling of the left and right 
spiral amplitudes 
 $via$
the forcing and azimuthal perturbation only if the frequencies $\omega_e$ 
and $\omega_h$
as well as the azimuthal wave numbers $m$ and $n$ are related as
\be
\frac{2\omega_h}{\omega_e}=k , \qquad \frac{2m}{n} = l, \qquad k,l=1,2,3...  
\ee
Simply put, under these conditions the term
$ B \,b^k \,w^l \,e^{iqz-i\omega_h t-im\phi} e^{i k \omega_e t} e^{i l n\phi} $
corresponds to the same Fourier coefficient in $z$, $t$ and $\phi$ as
$A e^{iqz+i\omega_h t+im\phi}$ and can 
therefore arise in the evolution equation for $A$, thus providing the crucial 
{\it linear} coupling. 
To cubic order one then obtains the following equations,
\bea
\partial_T A &= &a A + d b^k w^{2m/n} B + c|A|^2A + g|B|^2 A, \label{e:A}\\
\partial_T B& = &a^* B + d b^k w^{2m/n} A  + c^*|B|^2 B + g^*|A|^2B,\label{e:B}\\
\partial_T b &= &0, \qquad \partial_T w =0.\label{e:bw}
\eea
The real part $a_r$ of the coefficient $a$ gives the growth rate of the 
spiral waves and is proportional to the distance from the Hopf bifurcation, 
$a_r=\alpha (R_i-R_{ic})$, with $R_i$ the Reynolds number for the 
inner cylinder and $R_{ic}$ its critical value.
To allow for a small detuning in the forcing, the 
frequency $\omega_h$ in (~\ref{e:v}) is replaced by $\omega_e/2$. 
The detuning $\xi$ is then related to the 
imaginary part $a_i$ as
\be
\xi \equiv \frac{\omega_h-\omega_e/2}{\omega_h}=
\frac{a_i-\gamma a_r}{\omega_h}, 
\label{e:detun}
\ee 
where the term $\gamma a_r$ accounts for the change of the linear frequency 
with the control parameter $R_i$. The coefficients $d$, $c$ and $g$ are in 
general complex 
whereas
the product $db^kw^{2m/n}$ can be chosen real.
All coefficients are functions of the invariants $|b|^2$ and $|w|^2$.
This is most relevant for the small coefficient $a$. Thus the forcing and the 
azimuthal perturbation lead to a shift in the threshold and of the Hopf 
frequency even in the non-resonant case.

Eqs.(~\ref{e:A},~\ref{e:B}) show
that the forcing and the azimuthal perturbation have maximal effect if 
$k=1$ and $l\equiv 2m/n=1$, i.e.
if the external frequency is twice the Hopf frequency and the wave number
of the azimuthal perturbation is twice that of the spiral waves. 
Unfortunately, the latter
condition cannot be implemented easily, since eccentrically 
mounted cylinders
yield a perturbation of the form $\cos \phi$, i.e. $n=1$. 
 Clearly, {\em in the absence of any azimuthal perturbations ($w=0$) no 
linear coupling of oppositely traveling waves occurs and the standing
waves in question cannot arise directly from the basic state $A=0=B$
through a linear instability}.

The amplitude equations (~\ref{e:A},~\ref{e:B}) have been previously 
analyzed in detail 
\cite{RiCr88,Wa88} for the case $g_r <c_r<0$, i.e. for a Hopf bifurcation 
to stable
traveling waves rather than (unlocked) standing waves. 
The main result is the excitation of standing waves, $|A|=|B|=const$,
by the periodic forcing below the Hopf bifurcation ($a_r<0$). 
As is clear from the fact that
the amplitudes $A$ and $B$ are constant in time these waves are phase-locked 
to the forcing (cf. (\ref{e:v}) with $\omega_h \rightarrow \omega_e/2$).
A sketch of a typical phase diagram is shown in Fig. 1 where the loci of 
various transitions
are given as functions of the forcing strength (measured by $b$) and the mean 
Reynolds number (measured by $a_r$). 
The transition from the basic state of purely azimuthal flow to the 
phase-locked standing waves 
(ribbons) occurs along the line marked PSW, which is given by 
\be
b^2 = \frac{a_r^2+a_i^2}{(dw^{2m/n})^2}= 
\frac{a_r^2+(\xi \omega_h+\gamma a_r)^2}{(dw^{2m/n})^2}.\label{e:neutral}
\ee
Along the line marked H the basic state undergoes a 
Hopf bifurcation 
to traveling waves as well as (unstable) standing waves. 
This bifurcation exists 
already in the absence
of any periodic forcing ($b=0$). 
Along the line marked PB the phase-locked standing waves become unstable
to the traveling waves in a secondary parity-breaking bifurcation. Not 
relevant for this
experiment are the transition of the unstable phase-locked standing waves 
to unstable standing waves at SW and the saddle-node bifurcation of the 
phase-locked 
standing waves along SN.

To obtain quantitative
results pertaining to spiral vortex flow the linear as well as the nonlinear 
coefficients have to be determined numerically. This is a formidable task 
and will 
not be attempted here. It should be noted, however, that to leading order the 
nonlinear coefficients $c$ and $g$ are not affected by the forcing and the 
azimuthal
perturbation. 
Thus, the values for the concentric case can be used, 
which have been determined previously \cite{Demay}.

In addition to the results on modulated spirals in the slightly eccentric Taylor
system, we will also present results on
 traveling inclined rolls, which occur in the
Taylor-Dean system\cite{muta2}. In this system the azimuthal symmetry is 
broken strongly. Therefore, instead of the discussion presented above the 
original
derivation of the amplitude equations in the presence of a single translation 
symmetry applies \cite{RiCr88,Wa88,Ri90a}. It is based on an expansion of the 
flow
field in the form 
\be
{\bf v} =A(T) \,e^{iqz+i\omega_h t} \,{\bf f}_a(r,\phi)+B(T)\, e^{iqz-i\omega_h 
t} \,{\bf f}_b(r,\phi)+c.c.+h.o.t. \label{e:v2}
\ee
and leads to (\ref{e:A},\ref{e:B}) with $w=1$.

\section {Experimental Apparatus and Procedures} 

	Two variations of the concentric rotating cylinder (Taylor-Couette) 
system were employed to test the concepts of section 2.  
The systems break the rotational symmetry in very
different ways.  In the next few sections, we will 
describe each one in detail.  
First, we present the experimental system with strongly broken rotational 
symmetry, the Taylor-Dean system.

\subsection{ Taylor-Dean System}

\indent  The Taylor-Dean system consists of two independently rotating
horizontal coaxial cylinders with a partially filled gap (see Figure
2)\cite{muta1,muta2}.   
The partial filling of the
gap in the Taylor-Dean system breaks the rotational symmetry of the flow.  The
rotation of the cylinders and the two free surfaces impose a pressure gradient
along the azimuthal direction.  As a result, the flow sufficiently far away
from the free surfaces is a combination of Couette flow due to rotation of
the cylinders and Poiseuille flow due to the azimuthal pressure gradient.  
The main control parameters of this system are the inner and outer cylinder 
Reynolds numbers, $R_{i}=2\pi f_{i}r_{i}d/\nu$ and 
$R_{o}=2\pi f_{o}r_{o}d/\nu$, where $f_{o}$ and $f_{i}$ are the 
outer and inner cylinders' rotational frequencies, $d$ is the 
gap width, and $\nu$ is the kinematic viscosity of the fluid.  
 As the control parameters are varied the base
flow instabilities change from those associated with Taylor-Couette to
those associated with Dean flow.   The first experimental work with the 
Taylor-Dean system was done by Brewster and Nissan\cite{brew}, who 
studied the threshold of the
instability when the inner cylinder rotates and the outer cylinder is 
stationary.   More recently, Mutabazi, Normand, Peerhossaini,and 
Wesfreid\cite{muta3} have solved the linear stability problem for axisymmetric
and non-axisymmetric perturbations in the flow.  They found both 
stationary and traveling wave instabilities
depending on the ratio of the 
angular velocities $\mu$ ($=f_{o}/f_{i}$ )\cite{muta1}. 
We kept the outer cylinder stationary $(R_{o} = 0)$ for this 
study, while rotating the inner cylinder.   In this case the first
transition from the unperturbed base flow is to traveling inclined rolls 
as the inner cylinder speed increases, as shown in the experimental study of
Mutabazi, Hegseth, Andereck and Wesfreid\cite{muta2}.  
The transition to traveling inclined
rolls is, within the experimental error, a supercritical Hopf bifurcation.
At onset these rolls also have no preferred direction and may 
move either left or
right along the cylinder axis.  Therefore, this system can be used to verify
the theoretical prediction that a breaking of the time translational symmetry
by a small periodic modulation of the control parameter will result in a
stable standing wave pattern\cite{RiCr88,Wa88}.

\indent  Our experimental system consists of an inner cylinder made of black
Delrin plastic with radius $r_{i}=4.47 cm$ and a stationary outer cylinder 
made of Duran glass with radius $r_{o}=5.08 cm$, giving a gap 
$d = r_{o} - r_{i} = 0.59 cm$ and radius ratio $\eta = r_{i}/r_{o} = 0.883$. 
Two plastic rings are attached to the inner cylinder a distance 
$L = 52.4 cm$  apart, giving an aspect ratio $\Gamma = L/d = 88$,  large enough
to conceive of this as an extended system where one can  
neglect end effects. 
In this system the filling level fraction $ n = \theta_{f}/2\pi $ has been 
fixed at 0.75, where $\theta_{f}$ is the filling angle.   

The working fluid was pure double distilled water or a solution of 
double distilled water and 44$\%$ glycerol
by weight with 1$\%$ of Kalliroscope AQ1000 added for visualization. 
These nearly two-dimensional ($\approx 30\times 6\times 0.07 \mu m$) polymeric 
flakes align along the streamline
surfaces, reflecting light according to their orientation. Generally 
the dark areas
indicate flow along the observer's line of sight, while the light areas
indicate flow perpendicular to the line of sight.  
The apparatus was kept in a temperature controlled room so that
the temperature of the working fluid was held constant to within $0.1^0C$.

Spatial and temporal properties of the flow patterns were obtained using a
$512 \times 480$ pixel CCD camera connected to an image processor.  The image
processor board, installed in a PC, captured a picture of the flow pattern and
then a software routine was used to obtain the intensity along a single line
parallel to the axis of the cylinders.  This system is able to process up to
one line every $0.11\; sec$.  The data were then transferred to a VAX
4000-90 computer system and an analysis of the resulting intensity versus axial
position as a function of time plots (space-time data plots) yields the
wavelengths and the dynamics of the patterns in time and space. 

In the Taylor-Dean system, 
we have used a combination of two stepper motors 
(Compumotor A83-93) to drive the inner cylinder, one 
motor to produce a net
rotation and the other to produce a sinusoidal modulation of the inner
cylinder angular velocity.  The first motor was directly connected 
to the inner cylinder and
rotated with constant angular velocity.   The housing of this motor
was oscillated by a push-rod arrangement  mechanically coupled to the second
motor(see Figure 2).   The rotation of the second motor gave a net 
output at the inner cylinder of a constant angular velocity
plus a periodic sinusoidal variation in angular velocity.  

The motor speeds were controlled  through  Compumotor 2100 Series 
indexers 
and could be changed either manually or through computer control.  The motor
speeds have a frequency accuracy of 0.02$\%$.  Typically, a computer 
controlled 
the rotation speed, direction, and ramping rates.  Because both the
frequency and amplitude of the modulation could be varied
this introduced two new control parameters into the system, as
required by the theory.  

The Reynolds number of the inner cylinder is now 
$ R = R_{i} + R_{m} sin (2\pi f_{m} t)$, 
where $R_{m}=2\pi f_{a}r_{i}d/\nu$ and $f_{a}$ is the amplitude of the
sinusoidal rotational frequency. 
The two dimensionless parameters are
$ R_{m}/R_{ic}$ and the detuning 
parameter $\xi = (2f_{h} - f_{m})/2f_{h}$, where
$R_{ic}$ is the critical Reynolds number for onset of the traveling roll 
pattern and $f_{h}$ is the frequency of the teaveling rolls (Hopf
frequency) in the absence of modulation.  

To obtain the location of the onset 
of patterns for each modulation frequency $f_{m}$ and modulation amplitude 
$R_{m}$, we employed the following method.  
We first set the $R_{i}$ value below the 
onset of instabilities and then increased it quasistatically (keeping both
amplitude $R_{m}$ and frequency $f_{m}$ of modulation at a fixed value) until 
a flow pattern appeared in the system.  Then several sets of space-time data 
were taken while  increasing $R_{i}$. 
We repeated this procedure with increasing amplitudes and various frequencies
around twice the Hopf frequency (the frequency of the traveling rolls near
onset when there is no modulation). 

\subsection{ Results and Discussion}

\indent When there is no modulation $(R_{m}=0)$, the base flow 
bifurcates supercritically to a traveling roll pattern as we increase 
the inner cylinder speed.
A typical  space-time diagram for the traveling roll pattern near 
onset ($R_{i} = 265, R_{o}=0$, and $R_{m}=0$) is shown in Figure 3.  
The wavelength 
of the rolls along the cylinders is $\lambda = 0.841$ cm.  
At $\epsilon$ $(=\frac{R_{i} - R_{ic}}{R_{ic}})$ slightly greater than $0.0$
the pattern fills most of the working space and both left and
right traveling rolls may exist with a vertical defect line between them
(see Figure 4(a)).  
The frequency of the traveling rolls 
(Hopf frequency) near the onset ( $R_{ic} = 263$) is 0.543 Hz.    
Upon further increase of $R_{i}$, the flow undergoes a second instability 
to a short wavelength modulation of the traveling rolls at $R_{i}=303$, 
and then to an incoherent pattern at about $R_{i}=338$\cite{hegseth}.

The transition sequence changes dramatically with modulation of the inner
cylinder speed.  When we modulated the inner cylinder sinusoidally near
detuning parameter $\xi \approx 0$, we found standing waves rather than the 
traveling rolls (see Figure 4(b)), as 
predicted by theory.  
Figure 5 shows a
phase diagram of the primary transitions from the base flow to standing 
waves and the secondary transition from standing waves to 
traveling waves for $\xi \approx 0$ as we varied the modulation amplitude 
$R_{m}$ and inner cylinder Reynolds number $R_{i}$. 
One interesting feature to note in Figure 5 is that,
when the modulation amplitude is increased above a critical value 
($R_{m}/R_{i} > 0.05$),
the standing waves can be excited at $R_{i}$ values much lower than the critical
$R_{i}$. Another interesting feature is that as the amplitude of modulation
increases, we 
observe standing waves over a widening range of $R_{i}$.
The traveling wave state reappears when $R_{i}$ increases, as has been
observed in other systems\cite{rehberg} and in agreement 
with the theory\cite{RiCr88}.

At small modulation amplitudes 
($R_{m}/R_{i} < 0.05$) only the traveling roll state appears when $R_i$ is 
increased, also in agreement with the theoretical predictions (c.f. Figure 1). 

The space-time diagram
of the standing waves for $R_{m}/R_{ic} = 0.30$,
$R_{i} = 241$ and $\xi = 0 \pm 0.01$ is shown in Figure 6.  
During one modulation
period the light intensity at a given axial
position varies periodically, indicating the presence of
standing wave patterns. The frequency of the standing waves is half the 
modulation frequency.  

A quantitative analysis of the patterns was carried out using 
2-dimensional Fourier transforms of the patterns in time and space. 
This proved to be a very useful method for decomposing 
the standing wave patterns into their left and right traveling wave 
components from the original space-time CCD data.  From the 
decomposition we obtained 
the spatial wavelengths, temporal frequencies, and the amplitudes of each 
component.  Figures 7(a) 
and 7(b) shows the right and left traveling waves obtained from 
the space-time data shown in Figure 6.  The power spectra 
obtained using the 2-dimensional FFT for both right and left traveling waves 
are shown in Figures 7(c), and 7(d).  The left and right components have 
similar space and time characteristics.  The small differences in amplitudes of
the spectral peaks may be attributed to slightly nonuniform lighting
conditions.

At a still higher 
inner cylinder speed the standing waves lose their
stability to a traveling wave state.  This transition is supercritical within
our experimental resolution.  No mixed standing/traveling rolls states have
been seen.  The space-time 
diagram for the traveling wave state just above the onset 
at $R_{i} = 269$, $R_{m}/R_{ic} = 0.212$
is shown in Figure 8.  The
decomposition of this space-time data and associated power spectra 
are shown in Figures 9(a), (b), (c), and (d).   These figures
show that one traveling mode dominates over the other, much weaker,
components present in the system.  

\indent We also tested the sensitivity of 
the onset of the patterns, and which pattern appears first, 
to variation of the detuning parameter $\xi$ in this system.  
The result for $R_{m}/R_{i}=0.3$, shown in Figure 10, indicates 
that the onset of the patterns is very sensitive to the value of the detuning 
parameter $\xi$ for this system.  When $\xi \approx 0$ the standing wave 
pattern appears at $R_{i}$ well below $R_{ic}=263$. As we changed $\xi$ 
away from $\xi = 0$ (lower or higher modulation frequencies than
$2f_{h}$) the onset of the patterns was delayed considerably.  
In fact, at higher modulation frequencies ($\xi < -0.2$) patterns 
appeared only  
at {\em supercritical} $R_{i} (> R_{ic})$ values for $R_{m}/R_{ic} = 0.3$.   
Figure 10 also shows that, for fixed $R_{m}$, as we move away from $\xi = 0$ 
the characteristics of the patterns also change, so that traveling 
wave patterns appear at onset 
($\xi < -0.08$ or $\xi > 0.04$ for $R_{m}/R_{ic}=0.3$)  
rather than standing waves.  
It is possible that even for these large values of $|\xi|$ standing waves 
would reappear as the critical mode for higher fixed
$R_{m}/R_{ic}$ than 0.3.

Figure 11 shows the ratio of 
the fractional power under the left (the one induced by the modulation in this
case) traveling wave spectra to the total power, 
 as a function of $\xi$.  This characteristic resonance curve 
illustrates the disappearance of the 
left traveling wave as
the detuning parameter $\xi$ shifts away from 0.  The maximum fractional
power was a little less than the expected 0.5, which can be attributed 
to the non-ideal lighting conditions. 
We also analysed the frequencies of the two components as a function of $\xi$.
Figure 12 shows the frequency ratio as a
function of $\xi$.   This plot shows the standing wave region clearly, where 
the two
components lock to the same frequency over a range of $\xi$.  
The primary transition to standing waves and the secondary transition to
traveling waves in $R_{m}/R_{ic}$ vs $R_{i}$ parameter space are very
sensitive to $\xi$ for $\xi$ near zero.  Figures 13 (a) and (b) give  
the phase diagrams when $\xi$ is equal to $-0.03$ and $+0.03$, which show 
these phase changes clearly. A comparison of Fig. 13 and Fig. 5 shows that the
 minimal modulation amplitude for the excitation of standing waves
is actually lower for $\xi=0.03$ than for $\xi=0$. This can be ascribed to
the dependence of the linear frequency of the waves on $R_i$ as well as 
$|b|^2$ and $|w|^2$ as expressed in
(\ref{e:detun}); optimal resonance is obtained for $a_i=0$ rather than 
$\xi=0$ (cf. (\ref{e:neutral})).

 Further increases in $R_{i}$ beyond the
traveling wave state resulted in the appearance of short wavelength modulations
near $R_{i} \simeq 300$  and incoherent 
patterns at $R_{i} \simeq 340$ for $R_{m}/R_{ic}=0.25$ and $\xi = 0$.  These
patterns and their decomposed left and right traveling components are shown in
Figures 14 and 15.

\subsection{ Eccentric Taylor-Couette System}

\indent In the Taylor-Dean system, the
rotational symmetry is strongly broken by the air/liquid interfaces.  
We now consider a system with slightly broken rotational symmetry, which can 
be modeled as a perturbation on a system with two translational symmetries
(strictly speaking, a pure translation in space along $z$ and a rotational
symmetry along $\phi$).

The classical concentric Taylor-Couette system has translational symmetry in
the axial direction and rotational symmetry around the axis.  
As we discussed in section 2, for this system
it is necessary to apply periodic forcing as well as an azimuthal perturbation 
to linearly couple the left and right traveling
spirals to produce the standing waves.  We produced an azimuthal perturbation 
in this system by offsetting the axis of the inner cylinder relative to that of
the outer cylinder while maintaining the two axes parallel, thereby
making the system eccentric.  

The
experiment was performed in the region of ($R_{o},R_{i}$) parameter space where
traveling waves in the form of spiral vortices occur as the primary
instability. This happens when 
the cylinders are counter-rotating and concentric 
\cite{andereck,langford,tagg,tenna}. 
The spiral patterns 
travel in both the axial and azimuthal directions, and they break both the
axial and azimuthal symmetry of the base flow.  For a given radius
ratio there is a unique value of outer cylinder speed above which the primary
bifurcation from the base flow is a supercritical Hopf bifurcation 
to the time periodic spiral flow\cite{langford,tagg}.  The azimuthal wave 
number ($m$) of the spirals increases as the outer cylinder speed increases.  
The locations of the crossover points between spirals with
different azimuthal wave numbers are uniquely determined by the radius ratio of
the two cylinders\cite{tagg}.  

Our eccentric Taylor-Couette cylinder system, shown schematically 
in Figure 16, consists of an inner cylinder made of black Delrin plastic 
with radius $r_{i}=4.76 cm$, and 
an outer cylinder made of Plexiglas with radius $r_{o}=5.95 cm$, which gives 
a gap $d=r_{o}-r_{i}=1.19 cm$, and a radius ratio $\eta=0.800$.  The main 
control parameters are the inner and outer cylinder Reynolds numbers, 
$R_{i}=2\pi f_{i}r_{i}\bar{d}/\nu$ and
$R_{o}=2\pi f_{o}r_{o}\bar{d}/\nu$, 
where $\bar{d}$ is the average gap width.  A stationary support 
holds the apparatus at the top and the bottom.  The outer cylinder is connected 
to the stationary support by means of bearings at both ends.  
The upper end of the inner cylinder is attached to a long shaft which hangs 
from a horizontally movable plate on the stationary framework (see Figure 16). 
The lower end of the inner cylinder is left unattached, and is suspended 
$\approx 1 mm$ above the bottom of the system. 
 Eccentricity is adjusted by offsetting the axis of the inner cylinder relative
to the fixed axis of the outer cylinder.  The         
position of the axis of the inner cylinder is read to
an accuracy of $0.005 cm$ using a micrometer attached to the stationary support.
The eccentricity is $e=\varepsilon/(r_{o}-r_{i})$, where $\varepsilon$ 
is the offset of the two cylinder axes.  To maintain consistent 
end conditions, the upper and lower boundaries of 
the flow are formed by Teflon rings attached to the outer cylinder and
located near the ends of the cylinder.  There is a narrow gap of 0.4 cm 
between each of
these rings and the inner cylinder that is nevertheless wide enough to
allow for offsetting the inner cylinder. 
 The length of the fluid column $L$ is $40.40$ cm, giving an aspect ratio
$\Gamma=\frac{L}{r_{o}-r_{i}}=34.0$, large enough to minimize 
end effects.
Again, the working fluid was pure double distilled water or a solution of 
double distilled water and 44$\%$ glycerol
by weight with 1$\%$ of Kalliroscope AQ1000 added for visualization. 
The experimentally  obtained critical Reynolds number for concentric cylinders
when $R_{o}=0$ has been compared with reference values to check the 
accuracy of $\nu$ obtained from tabulated data\cite{crc}.  The difference was 
less than 1$\%$.
Both inner and outer cylinders are driven by two independently rotating 
Compumotor stepper motors, which are controlled through Compumotor 2100 Series
indexers.  Since the modulation amplitude required in the eccentric 
Taylor-Couette system was much greater than for the Taylor-Dean system, we 
could not use the two motor push rod arrangement to produce both linear and 
sinusoidal modulation of inner cylinder.  Therefore, we have used a single 
motor to
produce both a constant rotation speed component 
and the sinusoidal modulation of the inner 
cylinder rotation rate by sending 
control commands (a sequential change of shaft motion parameters with time
delays) to the indexer using the computer.  This produces an approximate
sinusoidal variation, with more than 40 step changes in velocity per cycle.

To obtain the location of the onset of patterns for each modulation frequency
$f_m$ and modulation amplitude $R_{m}$, we employed a method similar to that 
used in the experiment on the Taylor-Dean system.  We kept the outer cylinder 
speed fixed at a value within the range where the bifurcation to 
spiral vortices occurs.
Then, at a fixed modulation amplitude $R_{m}$ and constant frequency $f_{m}$,
we increased the average inner cylinder speed until the pattern appeared
for fixed modulation amplitude and constant frequency.  Then several sets of
space-time data were taken while increasing $R_{i}$.
We repeated this procedure with increasing amplitudes $R_m$ and various 
detuning parameters $\xi$ (i.e., frequencies
around twice the Hopf frequency, where the Hopf frequency in this case is 
the frequency of the spiral patterns near
onset in the absence of modulation). 

\subsection{Results and Discussion }

\indent When there is no modulation the base flow bifurcates supercritically to
spirals as we increase the inner cylinder speed while keeping the outer cylinder
speed constant.  A typical space-time diagram for the spiral pattern near onset
($R_{i}=151,R_{o}=155$, and $R_{m}=0$) is shown in Figure 17. 
Here the intensity of a line parallel with the axis of the cylinders was
recorded every $0.6s$ for $100s$.  The wavelength of the spirals along the axis
of the cylinder is $\lambda = 2.07cm \approx 2d $. 
Light sheet visualization through the gap cross section shows that the
spirals exist near the inner cylinder.   At 
$\epsilon$ $(=\frac{R_{i} - R_{ic}}{R_{ic}})$ slightly greater than zero
the pattern fills most of the working space along the axis. An upward 
moving spiral exists
near the bottom of the system and a downward moving spiral exists near the 
top. 
A horizontal defect line forms where the two spirals meet.  
These defect lines are not necessarily halfway between the top and the bottom
along the axis of the cylinders. 
Such defects are inherent to
traveling wave patterns\cite{gil}.  The frequency of the spirals (Hopf
frequency) with azimuthal wavenumber $m=2$, as shown in Figure 17, near 
the onset ($R_{ic}=151, R_{o}=155$ for $e=0$) is $0.113 Hz$. 

When we sinusoidally modulated the inner cylinder rotation speed, we
observed wave patterns which resemble standing waves (time-dependent 
patterns which are stationary in space)
rather than traveling waves (spirals) as the first
bifurcation from the base flow.  Figure 18 shows a space-time diagram of the 
standing-wave-like pattern near the
transition to the spiral state with azimuthal wave number $m=2$ at  
$R_{i}=151,\, R_{o} = 155, \,\frac{R_{m}}{R_{i}}=0.4, \, e=0.126$, and 
$\xi \approx 0$.  
Figure 19(a) 
and 19(b) shows the resulting right and left traveling waves obtained from 
the space-time data shown in Figure 18.  The power spectra 
obtained using the 2-dimensional FFT for right and left traveling wave
components, shown in Figures 19(c) and 19(d), have the 
same frequency characteristics, but different peak amplitudes.  A true standing
wave would have equal amplitudes in the traveling components. Even at very 
high modulation amplitudes $R_m$, one spiral
component always dominates the other.   Figure 20 shows the ratio of the
amplitudes of the primary peaks of the two spiral components versus modulation
frequency $f_{m}$.   

Figure 21 shows the phase diagram of $R_m/R_{ic}$ vs $R_i$ for $e=0.168$ 
and $\xi \approx 0$.  This shows that the base flow became unstable at lower
$R_{i}$ as the amplitude of modulation $R_m$ increased, in agreement with the 
theoretical 
predictions.  Also, the standing-wave-like patterns appeared only at large 
modulation amplitudes ( for $e=0.168$, $\frac{R_{m}}{R_{i}} > 0.17 $)
when the detuning parameter $\xi \approx 0$.  
At small modulation amplitudes 
($R_{m}/R_{i} < 0.17$) only the traveling roll (spiral) 
state appears when $R_i$ is increased.  Whether standing-wave-like patterns 
are produced also depends on the eccentricity of the system.
For small eccentricity $e$, they appeared only at higher 
modulation amplitudes, in agreement with the theoretical predictions. 

Compared with other systems (e.g., the Taylor-Dean system discussed in the
previous section or electroconvection of nematic liquid crystals and binary 
fluid convection \cite{rehberg}), 
in this system, the standing-wave-like patterns with two components having 
the same frequency appeared only in a small region of parameter space near the 
resonance frequency (see Figure 22).  Both the small region of parameter space
and the unequal power spectra amplitudes indicate that the coupling of the 
modulation to the fluid flow is 
very weak.  Two factors may contribute to this.  Eqs.(~\ref{e:A},~\ref{e:B}) 
show that the azimuthal symmetry
breaking employed in the experiment ($n=1$) 
enters the coupling between left- and right-traveling spirals only
with its fourth power for spirals with $m=2$.  Thus, quite large eccentricities 
are needed to give a strong effect. Motivated by this observation we have also 
studied the modulation of $m=1$-spirals.  However, no resonant excitation of 
standing waves was observed; instead, the patterns relax to the base flow during
part of the cycle.  Since the frequency of the waves is very small,
they appear to follow the periodic forcing adiabatically indicating that their
growth rate is of the same order as the forcing frequency.  Therefore for 
(~\ref{e:A}, ~\ref{e:B}) to apply,
the growth rate would have to be reduced substantially.  This would require
much higher resolution in the average Reynolds numbers than is accessible
in the present apparatus.  The investigation of $m=1$-spirals is further 
complicated by the presence of other modes like axisymmetric vortices
and interpenetrating spirals at nearby parameter values.

A second reason for the apparent weakness of the  
coupling is presumably the penetration depth of the
oscillation into the bulk of the fluid, as characterized by the 
viscous Stokes layer of width 
$\delta \, = \, \sqrt{\nu/\omega}$, near the
inner cylinder.  The thickness of the Stoke's layer $\delta$ 
at $\xi = 0$ for both $m=1$ and $2$ spiral states are $0.24cm$ and $0.17cm$, 
much smaller than the size of the gap $d = 1.19cm$.  
Thus, the bulk of the fluid feels the 
modulation only very weakly. (It is possible that similar effects may have been
present in the binary fluid convection experiment reported in \cite{rehberg}. 
This may have contributed to the less well-defined standing waves reported
there, in contrast with the electroconvection experiments.)  
Futhermore, in the eccentric
Taylor-Couette system, modulation creates only a weak azimuthal pressure
gradient due to the small eccentricity, and the oscillatory Stokes layer
influence may not be enough to excite a strong secondary traveling component.
In the Taylor-Dean system, on the other hand, the inner cylinder modulation
creates not only the oscillatory boundary layer influence but also an 
induced oscillatory azimuthal pressure gradient across the whole gap, due to 
the free surface, which more readily affects the pattern.

The standing-wave-like patterns appearing in the eccentric Taylor-Couette
system at the $\xi=0$ near the transition
boundary to $m=2$ spiral state lose their stability to an 
incoherent pattern when the inner cylinder speed is increased further 
above the onset.  This may be due to the loss of stability of the basic 
spirals to interpenetrating spirals and
wavy spirals.   Even for the concentric cylinder case ($e=0$) we have 
seen some evidence of a weak secondary spiral at very high 
modulation amplitudes.  This could be due to small imperfections in 
the experimental apparatus (e.g. nonparallel cylinder axes) which
may break the rotational symmetry sufficiently to induce a second spiral.  

\section{ Conclusions }
\indent In conclusion, we have found that time-periodic 
modulation at close to twice the frequency 
of a Hopf bifurcation can induce standing waves in
systems with traveling wave patterns in two directions. The Taylor-Couette
system with counter-rotating cylinders produces spirals at onset over a large
parameter range, but unless the rotational symmetry of the apparatus is 
broken standing waves cannot be excited by modulation.  
This is in contrast with the case of one-dimensional traveling wave patterns
studied in convection\cite{rehberg},
where only temporal modulation is necessary to excite standing wave
patterns.  Temporal modulation of the eccentric Taylor-Couette system induces 
a second traveling spiral pattern as predicted by theory. However, owing 
partly to weak coupling of the oscillations to the bulk of the
flow, it never grows to a large enough amplitude, nor does it
couple with the other traveling spiral sufficiently strongly, to produce simple 
standing wave patterns at onset.  The temporally modulated Taylor-Dean system,
in contrast, with its strongly broken azimuthal symmetry, yields clear agreement
with the qualitative features of the theoretical model.

\section{Acknowledgements} 
\indent  CDA, SGKT, and JJH thank ONR for support through grants
N00014-86-K-0071 and N00014-89-J-1352.  HR gratefully acknowledges 
discussions with J.D. Crawford and E. Knobloch. 
His work was supported by DOE through grant DE-FG02-92ER14303.

\newpage

{\large \bf Figure Captions}

\normalsize

\noindent Figure 1: A sketch of a typical phase diagram for modulated
rotation at resonance ($\omega_{e} = 2 \omega_{h}$), where the loci of various
transitions are given as a function of the forcing strength (measured by $b$) 
and mean Reynolds number (measured by $a_{r}$). 
The transition from the basic state to standing waves occurs along the line 
marked PSW and from the basic state to traveling waves occurs along the 
line marked H.

\noindent Figure 2: (a) Schematic diagram of the Taylor-Dean apparatus. (b) 
Schematic cross section of the apparatus. The front side is the 
side where the observer sees
the inner cylinder rotating upward as shown.

\noindent Figure 3: Space-time diagram of the traveling roll pattern
in the Taylor-Dean system when
$R_{i}=265,\; R_{o}=0$, and $R_{m}=0$.  The wavelength of the pattern is 
$\lambda = .841 cm$ and frequency = $0.543 $ Hz.

\noindent Figure 4: Pictures of the (a) traveling roll (for $R_{i}=269$ and 
$R_{m}=0$) and 
(b) standing wave states in the Taylor-Dean system (for $R_{i}=265$ and 
$R_{m}/R_{ic}=0.28$).

\noindent Figure 5: Phase diagram of standing wave and traveling wave states of
the Taylor-Dean system when the detuning parameter is $\xi = 0$. 
The plus and cross signs indicate the location of the bifurcation 
from the base state to  
standing waves and standing waves to traveling waves respectively.  

\noindent Figure 6: Space-time diagram of the standing wave state at 
$R_{m}/R_{i} = 0.30$, $R_{i} = 241$ and $\xi = 0 \pm 0.01$ (Taylor-Dean
system).

\noindent Figure 7: Space-time diagrams of (a) the right traveling wave
component and (b) the left traveling wave component of the standing wave state 
shown in Figure 6, obtained using a 2-dimensional FFT decomposition.  
(c) and (d) are the frequency power spectra of the
decomposed right and left traveling modes. The frequencies
are scaled to the inverse of the diffusive time scale $\nu/d^2$.

\noindent Figure 8: Space-time diagram of the traveling wave state at
$R_{m}/R_{ic}=0.212,\; \xi=0.0 $, and $\; R_{i}=269$ just above 
the onset (Taylor-Dean system). 

\noindent Figure 9: Space-time diagrams of (a) the right traveling wave
component and (b) the left traveling wave component of the traveling wave state 
shown in Figure 8.  (c) and (d) are the frequency power spectra of the
decomposed right and left traveling modes. The frequencies
are scaled to the inverse of the diffusive time scale $\nu/d^2$.
The right traveling wave component dominates the left traveling component 
in the system.

\noindent Figure 10: The onset of the primary flow transition $Ri$ vs the 
detuning parameter $\xi$  $(= (2f_{h} - f_{m})/2f_{h})$ at 
$R_{m}/R_{ic} = 0.3$ (Taylor-Dean system).  
 
\noindent Figure 11: The fractional power of the secondary traveling wave state
vs the detuning parameter $\xi$ at $R_{m}/R_{ic} = 0.3$.

\noindent Figure 12: The frequency ratio of the two traveling wave components vs
the detuning parameter $\xi$ at $R_{m}/R_{ic} = 0.3$.  Both left and 
right traveling components have
the same frequency within experimental error over a fairly wide region near 
resonance $\xi = 0$.

\noindent Figure 13: Phase diagrams when (a) $\xi = -0.03$ and (b)
$\xi = 0.03$.   Pluses  and crosses indicate the transitions from
the base flow to standing waves and to traveling waves respectively.

\noindent Figure 14: (a) The traveling wave state with the short wavelength
modulations at $R_{i} = 300$ and $R_{m}/R_{ic} = 0.254$ (Taylor-Dean system). 
(b) and (c) are the 
decomposed space-time data 
for right and left traveling modes, respectively, of the same 
traveling wave state shown in (a). 

\noindent Figure 15: (a) The incoherent patterns at 
$R_{i} = 340$ and $R_{m}/R_{ic} = 0.254$ (Taylor-Dean system). (b) 
and (c) are the decomposed 
space-time data for right and 
left traveling modes, respectively, of the incoherent pattern shown in (a). 

\noindent Figure 16: (a) Vertical, and (b) horizontal, cross-sections of the
eccentric cylinders system.

\noindent Figure 17: Space-time diagram of the spiral pattern ($m=2$) in the
Taylor-Couette system at $R_{i} = 151,\; R_{o} = 155,\; e = 0$, and 
$R_{m}= 0$.

\noindent Figure 18: Space-time diagram of the standing-wave-like pattern in the
Taylor-Couette system at $R_{i} = 151,\; R_{o} = 155,\; e = .126$, 
$R_{m}/R_{ic} = 0.4$, and $\xi=0$. 

\noindent Figure 19: (a) and (b) are the decomposed components of the 
space-time data in Figure 18.   (c) and (d) are the frequency power spectra of 
the right and left traveling modes. The frequencies
are scaled to the inverse of the diffusive time scale $\nu/d^2$.
The right traveling wave component dominates the left traveling component 
in the system.

\noindent Figure 20: The ratio of the spectral peak amplitude of the secondary 
spiral wave state to the spectral peak amplitude of the primary spiral wave 
state vs the detuning parameter $\xi$ at $R_{i}=151,\; R_{o}=155,\;
R_{m}/R_{ic}=0.53$, and $e=0.126$.

\noindent Figure 21: Phase diagram of $R_m/R_{ic}$ versus $R_i$ for the 
eccentric Taylor-Couette system with eccentricity 
$e=0.168$ and detuning parameter $\xi = 0$. 

\noindent Figure 22: The frequency ratio of the two traveling wave components 
versus the detuning parameter $\xi$ in the eccentric Taylor-Couette system
at $R_{i}=151,\; R_{o}=155,\; R_{m}/R_{ic}=0.53$, and $e=0.126$.  
 The left and right traveling components have the 
same frequency only in a very small region near resonance $\xi = 0$.

\end{document}